\documentclass[doublecol]{epl2}

\title{Strong evidences for a nonextensive behavior of the rotation period in Open Clusters}
\shorttitle{Nonextensive behavior of the rotation period in Open Clusters}

\author{D. B. de Freitas$^{1}$\thanks{E-mail: \email{danielbrito@dfte.ufrn.br}}
\and M. M. F. Nepomuceno$^{1}$
\and B. B. Soares$^{2}$
\and J. R. P. Silva$^{2}$}
\shortauthor{D. B. de Freitas \etal}

\institute{
  \inst{1} Departamento de F\'{\i}sica,
    Universidade Federal do Rio
    Grande do Norte, 59072-970
    Natal,  RN, Brazil\\
\inst{2} Departamento de F\'{i}sica, Universidade do Estado do Rio Grande do Norte, Mossor\'o--RN, Brazil\\
}

\abstract{
Time-dependent nonextensivity in a stellar astrophysical scenario combines  nonextensive entropic indices $q_{K}$ derived from the modified Kawaler's parametrization, and $q$, obtained from rotational velocity distribution. These $q$'s are related through a heuristic single relation given by $q\approx q_{0}(1-\Delta t/q_{K})$, where $t$ is the cluster age. In a nonextensive scenario, these indices are quantities that measure the degree of nonextensivity present in the system. Recent studies reveal that the index $q$ is correlated to the formation rate of high-energy tails present in the distribution of rotation velocity. On the other hand, the index $q_{K}$ is determined by the stellar rotation-age relationship. This depends on the magnetic field configuration through the expression $q_{K}=1+4aN/3$, where $a$ and $N$ denote the saturation level of the star magnetic field and  its topology, respectively. In the present study, we show that the connection $q-q_{K}$ is also consistent with 548 rotation period data for single main-sequence stars in 11 Open Clusters  aged less than 1 Gyr. The value of $q_{K}\sim$ 2.5 from our unsaturated model shows that the mean  magnetic field topology of these stars is slightly more complex than a purely radial field. Our results also suggest that stellar rotational braking behavior affects the degree of anti-correlation between $q$ and cluster age $t$. Finally, we suggest that stellar magnetic braking can be scaled by the entropic index $q$.}

\begin{document}

\maketitle

\section{INTRODUCTION}
Open Clusters (hereafter OC) are particularly important given that their members have similar age and chemical abundance, and are physically related by their mutual gravitational attraction. The advantage of studying stars that are OC members, as opposed to field stars, has been recognized and exploited since the dawn of stellar rotation (e.g., \cite{Soares06,carvalho2007,carvalho2008,carvalho2009}). In particular, the pioneering study by Barnes \cite{barnes2003} was the first to analyze slow and fast rotator populations in OCs as distinct ``rotational sequences''. A number of authors 
(e.g.,\cite{kawaler1988,chaboyer1995}) have reported that the presence of these two rotator populations in field stars is associated with the different dependencies that link angular momentum loss rates to angular velocity. Kawaler \cite{kawaler1988} developed a parametrization for main-sequence stars showing that the rate of angular momentum loss is a power law in $\Omega$, where $\Omega$ denotes angular velocity.  This author also showed that for rotating stars as a rigid body, the time dependency of equatorial rotation velocity is deterministic and not merely a statistical artifact. This conclusion is also valid for differentially rotating stars, as reported by Krishnamurthi et al. \cite{kris1997}. This deterministic characteristic of stellar rotational evolution was recognized as early as Schatzman \cite{schatzman}, Kraft \cite{kraft62} and Skumanich \cite{Skumanich}. Other recent studies, such as Irwin \& Bouvier \cite{irwin2008}, Barnes \& Kim \cite{barnes2010}, Reiners \& Mohanty \cite{reiners2012} and de Freitas \& De Medeiros \cite{defreitas2013}, have also  confirmed that rotation data for cluster and field stars at different evolutionary stages and in different environments are deterministic.

The theory developed by Chaboyer et al. \cite{chaboyer1995}, based on Kawaler's parametrization, emphasized that the rate of angular momentum loss as a function of $\Omega$ is divided into two magnetic field regimes: one saturated at high rotation rates and the other unsaturated at slow rotation rates. According to Bouvier, Forestini \& Allain \cite{bouvier1997}, the saturated level for solar-type stars appears in the pre-main sequence phase, where the memory of initial angular momentum is retained up to a few 10$^{8}$ yr. On the other hand, as reported by de Freitas \& De Medeiros \cite{defreitas2013}, for lower mass stars the retention of this memory occurs up to an age of about 1 Gyr. Beyond these ages, stellar rotational velocity decreases as a power-law in $1+4aN/3$, where $a$ and $N$ denote the saturation level of the stellar magnetic field and its topology, respectively. As reported by Chaboyer \cite{chaboyer1995}, the $a$-parameter varies between 1 and 2 for the unsaturated regime and is 0 for the saturated regime, while $N$ ranges  from $\sim$ 0 to 2, where 0 denotes dipole magnetic field geometry and the other extreme indicates a purely radial field. However, these values of $a$ and $N$ are not entirely conclusive, as the original Kawaler \cite{kawaler1988} formulation for the angular momentum loss was merely an attempt to identify stellar parameters relevant to the problem of rotational evolution and their scaling properties. As such, the meaning of these parameters should not be taken too literally. In this context, the work by Silva et al. \cite{Silvaetal13} suggests that the memory of initial angular momentum can be scaled by the entropic index $q$ derived from rotational velocities. The authors proposes a connection between $q$ and the parameter $q_K$ from the nonextensive model created by de Freitas \& De Medeiros \cite{defreitas2013}. This nonextensive formalism describes stellar rotational evolution in the saturated and unsaturated regimes of magnetic fields based on the modified Kawaler's parametrization \cite{kawaler1988,chaboyer1995}. The present study proposes that the two regimes are connected by a nonextensive parameter denoted by $q_{K}$ (see footnote\footnote{The subscript $K$ stands for Kawaler.}) (for a review about the nonextensive theory, see references \cite{tsallis1988,tsallis2004}). For exponential decay, we have the saturated regime, i.e., constant magnetic field, while for the unsaturated regime ($q_{K}>1$) the magnetic field follows a power-law, like $\Omega^{a}$. In particular, a Boltzmannian system explains the behavior of saturated stars, whereas a complex (nonextensive) system defines the evolution of unsaturated stars.

As stated by Wright et al. \cite{wright2011}, in the unsaturated regime there are two influences on the efficiency of the magnetic dynamo: the stellar rotation period $P$ and mass-dependent convective turnover time $\tau$. Noyes et al. \cite{noyes1984} combined these two parameters into a single parameter known as the Rossby number $R_{0}=P/\tau$. The main goal of this relationship is to convert observable quantities into parameters of the stellar dynamo (in the present case, $P$ leads to $R_{0}$). Several authors (e.g., \cite{randich2000}) have shown that $R_{0}$ is an effective parameter of the stellar magnetic dynamo. On the other hand, Wright et al. \cite{wright2011} combine photometric rotation periods and X-ray luminosities $L_{X}$ to study and characterize the rotation--activity relationship as proof of  dynamo efficiency in the unsaturated regime. Moreover, Pace \& Pasquini \cite{pace2004} provided evidence for an age-activity relationship in solar-type stars of 5 OCs. These authors concluded that the evolution of chromospheric activity and rotation with age are virtually equal, exhibiting the same decaying trend. More recently, de Freitas \& De Medeiros \cite{defreitas2013} revisited the age-activity-rotation relationship problem. They suggest that the behavior of the $q_{K}$-index provides relevant information concerning  the level of magnetic braking in F- and G- type main sequence field stars older than 1 Gyr.

In this paper we present a nonextensive insight to investigate the behavior of rotational evolution from a sample of 548 rotation period data items for solar-type stars from 11 young OCs aged between 35 and 794 Myr. In particular, we revisit the connection $q$-$q_{K}$ proposed by Silva et al. \cite{Silvaetal13} and compare our results with those obtained using $V \sin i$ data. The structure of the present study is as follows: in the next section we explain our working sample and best-fit parameters; the third section presents results and discussion based on the physical implications of nonextensive indices and; finally, conclusions are given in the last section.

\begin{table*}
\scriptsize
\caption{Main characteristics of the data analyzed in this study. Column (1) identify the cluster. Column (2), and (3) present the references for rotation period, and the number of stars analyzed, respectively. Columns (4) shows the cluster ages, in Myrs, and Column (5) displays the median of the rotation period, in days. The best-fit parameters are given in columns (6) and (7), respectively. Column (8) gives the probability value from the Anderson-Darling test for the best-fit, and Column (9) presents the probability from the Anderson-Darling test comparing the best-fit curve and the KDE for each ECDF.}
\label{tab1}
\renewcommand{\arraystretch}{1.5}
\begin{tabular}{rc|rrr|ccr|r}
\hline \hline
(1) &(2) & (3) &(4) & (5) & (6)&(7) & (8)& (9)\\
Cluster & Ref. & $N$ & Age & $P^{-1}_{rot}$ & $q$ & $\sigma$ & $Prob_{\rm{f}}$ & $Prob_{\rm{K}}$\\
\hline \hline

$\alpha$Per & 1 & 28 & 35 & 0.3 & 1.56 $^{+ 0.050 }_{- 0.103 }$ & 3.03 $^{+ 3.322 }_{- 1.533 }$ & 0.464 &
0.470 \\[0.6ex]
IC\,2391 & 1 & 24 & 76 & 0.9 & 1.48 $^{+ 0.083 }_{- 0.148 }$ & 4.47 $^{+ 3.633 }_{- 2.080 }$ & 0.683 & 0.655
\\[0.6ex]
NGC\,2516 & 1 & 34 & 120 & 2.6 & 1.07 $^{+ 0.188 }_{(-0.115)}$ & 6.24 $^{+ 1.631 }_{- 1.450 }$ & 0.590 & 0.517
\\[1.5ex]
Pleiades & 2 & 97 & 120 & 2.6 & 1.27 $^{+ 0.063 }_{- 0.068 }$ & 4.58 $^{+ 0.551 }_{- 0.519 }$ & 0.141 & 0.165
\\[0.6ex]
M\,50 & 3 & 110 & 135 & 2.6 & 1.44 $^{+ 0.043 }_{- 0.043 }$ & 2.47 $^{+ 0.443 }_{- 0.369 }$ & 0.215 & 0.403
\\[0.6ex]
M\,35 & 4 & 112 & 151 & 2.6 & 1.41 $^{+ 0.059 }_{- 0.070 }$ & 3.32 $^{+ 0.746 }_{- 0.640 }$ & 0.299 & 0.375
\\[1.5ex]
M\,34 & 1 & 14 & 236 & 1.0 & 1.33 $^{+ 0.224 }_{- 0.276 }$ & 3.01 $^{+ 3.561 }_{- 1.156 }$ & 0.491 & 0.506
\\[0.6ex]
M\,37 & 1 & 61 & 347 & 4.0 & 0.96 $^{+ 0.198 }_{(-0.001)}$ & 4.60 $^{+0.926 }_{(-0.100)}$ & 0.400 & 0.424
\\[0.6ex]
NGC\,6811 & 5 & 37 & 575 & 5.7 & 1.10 $^{+ 0.140 }_{- 0.080 }$ & 2.65 $^{+ 0.550 }_{- 0.240 }$ & 0.110 & 0.264
\\[1.5ex]
Hyades & 1 & 19 & 794 & 8.3 & 0.94 $^{+ 0.250 }_{(-0.001)}$ & 2.67 $^{+ 0.760 }_{(-0.038) }$ & 0.443 & 0.377
\\[0.6ex]
Praesepe & 6 & 12 & 794 & 6.3 & 0.66 $^{+ 0.375 }_{(-0.038)}$ & 4.13 $^{+ 2.574 }_{(-0.230)}$ & 0.600 & 0.526
\\
\hline
\end{tabular}
\end{table*}

\section{WORKING SAMPLE AND BEST-FIT PARAMETERS}
In this study we analyzed 548 rotation period data items from single main-sequence stars in 11 open clusters aged between 35 and 794 Myr. The range of stellar masses is $0.9-1.1 M_{\odot}$, except for clusters NGC 6811, Hyades, and Praesepe, whose mass intervals are $1.1-1.4 M_{\odot}$, $0.7-1.1 M_{\odot}$, and $0.9-1.4 M_{\odot}$, respectively. \textit{Our main goal is to investigate the behavior of the nonextensive entropic index $q$ and its effects on rotational evolution after the period of gravitational contraction (Pre-Main-Sequence, PMS).} In this respect, the present paper deals with OC stars past their Zero Age Main Sequence (ZAMS). It is also worth mentioning that a sample of rotation periods is free of ambiguities arising from $\sin i$, contrary to what occurs with a sample of projected rotation (e.g., \cite{Silvaetal13}). Once the star-disk has disappeared in $\sim$5 Myr for low-mass stars, wind braking is the dominant process to counteract PMS contraction and later on for main-sequence spin down \cite{kawaler1988}. Thus, our working sample can be considered to be in the unsaturated regime, indicating that the computation of the values of the saturation parameter is not necessary.

In the present study, the true stellar equatorial rotation $V$ of the generalized distribution function (see Equation 1 from \cite{Silvaetal13}) is replaced by $P^{-1}$, where $P$ is the rotation period. Thus, we can rewrite the nonextensive probability distribution function as
\begin{equation}
\label{dis4}
f_{q}(P)=\frac{C}{P^{2}}\left[1-(1-q)\frac{\sigma_{P}^{2}}{P^{2}}\right]^{\frac{1}{1-q}}, \quad -\infty<q\leq 3.
\end{equation}
where $C$ is a $q$-dependent constant which should be calculated from the normalization of eq. (\ref{dis4}). For each $q<3$, there is a distribution $f_{q}(P)$ that maximizes the nonextensive entropy $S_{q}$, i.e., the distribution is normalized for $q<3$. If $q\geq 3$, the norm constraint cannot be satisfied. However, in order to avoid biases in the frequency histograms due to arbitrary choices of bin ranges, we used the empirical cumulative distribution function (ECDF) of $f_{q}(P)$ and compare it with the distribution function in eq. (\ref{dis4}), namely

\begin{equation}
\label{dis5}
F_{q}(P)=\frac{\int^{P}_{P_{min}}\frac{1}{P^{2}}\left[1-(1-q)\frac{\sigma_{P}^{2}}{P^{2}}\right]^{\frac{1}{1-q}}\mathrm dP}{\int^{\infty}_{P_{min}}\frac{1}{P^{2}}\left[1-(1-q)\frac{\sigma_{P}^{2}}{P^{2}}\right]^{\frac{1}{1-q}}\mathrm dP}.
\end{equation}
when $q<1$ it is necessary
to impose a restriction
on the $P$ values in order to ensure the positivity of $f_{q}(P)$ as
\begin{equation}
 P_{min}= \sigma\sqrt{1-q},
 \label{cutx}
\end{equation}
for $q\geq1$, $P_{min}=0$.

From nonextensive formalism, the index $q$ measures the degree of nonextensivity of the entropy in a system composed of several subsystems. In the context of stellar rotation, $q$ is related to the magnetic braking efficiency as shown by \cite{defreitas2013} and \cite{Silvaetal13}. Moreover, these authors also showed that the memory of initial angular momentum can be quantified using the index $q$, i.e., this efficiency can be scaled by $q$. 

In the nonextensive theory, for the particular case of two independent subsystems $A$ and $B$, the total entropy satisfies the nonadditive relation
\begin{eqnarray}
\label{qentro}
S_{q}(A+B)=S_{q}(A)+S_{q}(B)+\frac{1-q}{k} S_{q}(A) S_{q}(B),
\end{eqnarray}
where the entropic additive form of the Boltzmann-Gibbs entropy is recovered when $q=1$. 

The cases $q<1$ and $q>1$ correspond respectively to the superextensive and subextensive regimes of entropy $S_{q}$, assuming in all cases $S_{q}\geq0$. The $q$-distributions have tails that decay on the order of $q$ when $1<q<3$ and have bounded support when $q<1$. Non-equilibrium systems are deeply related to heavy-tailed distributions whose tails are not exponentially bounded. In general, these tails for $q>1$ are associated with the hierarchical structure in phase space that is non-uniformly occupied, in contrast to the expected behavior in equilibrium systems where B--G framework is valid. Specifically, for $q<1$ the formalism imposes a high-energy cutoff.

\begin{table}
\scriptsize
\caption{Best-fit values from the nonextensive model by de Freitas \& De Medeiros \cite{defreitas2013} for the time-dependence of the mean rotation period as displayed in Figure \ref{qkxages}.}
\label{tab2}
\renewcommand{\arraystretch}{1.5}
\begin{tabular}{ccccc}
\hline \hline
$q_K$ & $\lambda_{q_{K}}$ & $LogLik$ & Behavior of the model \\
\hline \hline
$1^{a}$ & $19.8\pm2.8$ & -29.6 & Exponential decay\\
$3^{a}$ & $124.9\pm35.4$ & -26.6 & Skumanich law\\
$2.5\pm0.65$ & $69.7\pm50.9$ & -26.3 & Unsaturated\\
\hline
\end{tabular}
\\
\end{table}

\begin{figure}
\begin{center}
\includegraphics[width=0.45\textwidth]{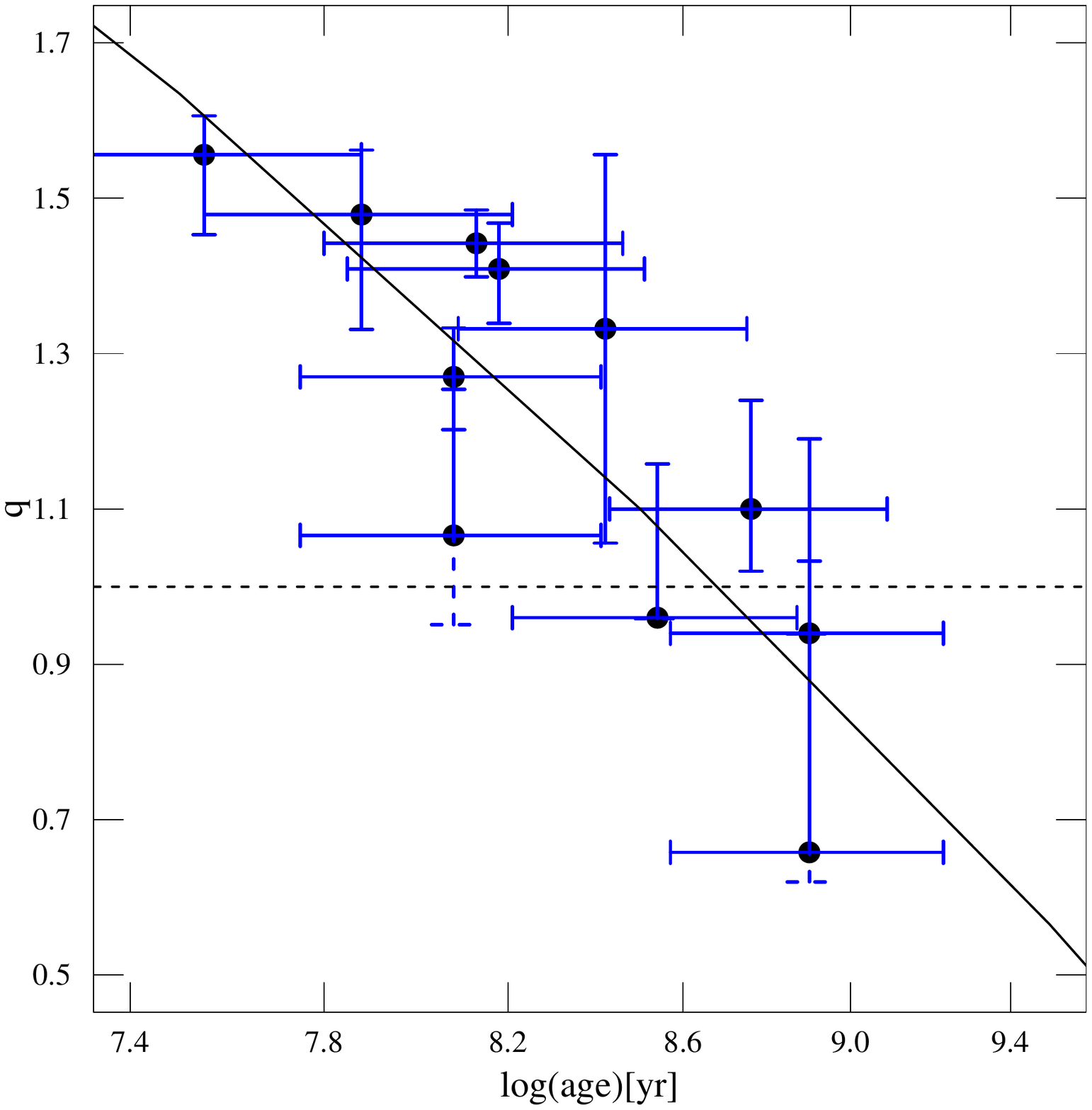}
\end{center}
\caption{Distribution of $q$ as a function of the cluster age. The vertical error bars represent the 95\% confidence interval, except the dashed ones which represent the minimum limit of $q$ which ensures the positivity of the distribution function. The continuous and dashed lines are the best-fit curve $q=-0.53\log (age) +5.64$, and the curve $q=1$, respectively. The point where these two lines intersect corresponds to $\log (age)=8.75$ dex (570 Myr).}
\label{qxages}
\end{figure}

\begin{figure}
\begin{center}
\includegraphics[width=0.45\textwidth]{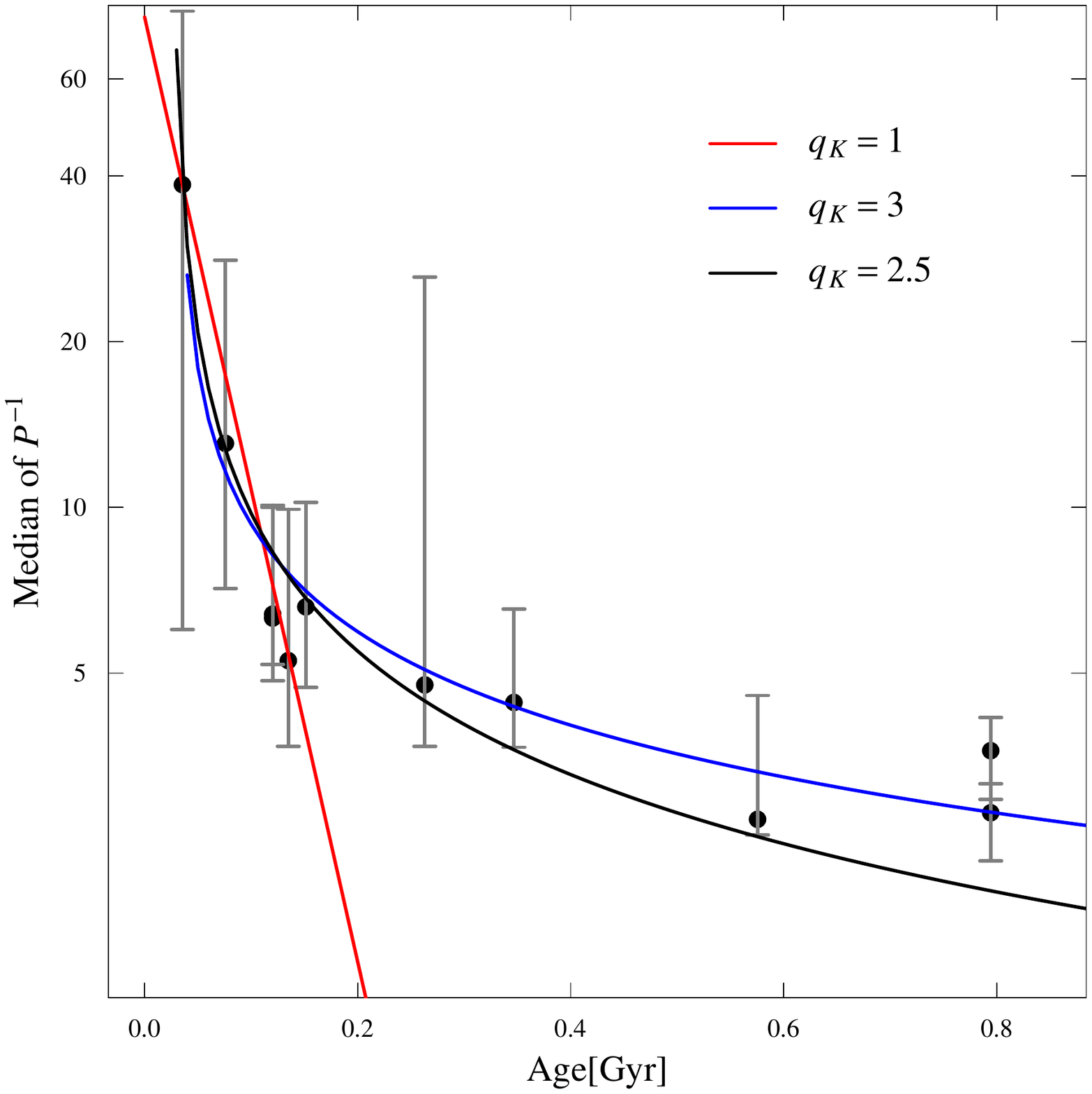}
\end{center}
\caption{Median of the inverse rotation periods, in solar unit, as a function of the cluster ages. Continuous lines represent the best-fit curves of the model proposed by \cite{defreitas2013}, and error bars correspond to the $5^{th}$, and the $75^{th}$ percentiles of the inverse rotation periods.}

\label{qkxages}
\end{figure}

The cluster ages were obtained from the Catalogue of Open Cluster Data (COCD) by Kharchenko et al. \cite{Kharchenko05}, except Hyades, whose age was obtained from the WEBDA database \cite{Mermilliod04}. Ages from the COCD were determined using the turn-off isochrone technique. The accuracy of this method is limited by the lack of high mass stars evolving off the main-sequence, particularly for young clusters. We used COCD data because it is a uniform database on cluster ages that uses a homogeneous cluster age scale. This is an important aspect, since we are analyzing the relationship between $q$ and stellar age, where cluster age is the chronometer. In order to account for the possible range of cluster ages, we used the 0.33 dex range adopted by Silva et al. \cite{Silvaetal13} as the possible range of individual cluster ages instead of the estimated value reported in Kharchenko et al. \cite{Kharchenko05} (i.e., 0.2--0.25 dex). In the next section we discuss the implications of adopting a wider range of cluster ages.
Best-fit parameters $q$ and $\sigma$ for the clusters were obtained from the ECDF of $x=1/P$, where $P$ is the stellar rotation period in solar units, assuming $P_{\odot}=26.1$ days (in \cite{Donahueetal96}, see Table 1). For a cluster with $n$ discrete data $x_i$ $(i=1,2,...,n)$ sorted in ascending order, the ECDFs are distributions that assign probability mass $1/n$ to each value of $x_i$, i.e.,
\begin{equation}
\label{pmf}
f(x)=\frac{1}{n}, ~~~~ x\in\left\{x_1, x_2, ..., x_n\right\}.
\end{equation}
When a value of $x_i$ is repeated $k$ times, the probability mass of $x_i$ is accumulated to $k/n$ \cite{qin98,kerns10}. We estimated the best-fit $q$ and $\sigma$ by numerically evaluating the integral of Equation (\ref{dis4}) to find the best-fit curve matching Equation (\ref{dis5}), according to the Anderson-Darling goodness-of-fit test \cite{Anderson54}. The best-fit parameters and probabilities obtained in the goodness-of-fit test ($Prob_f$) are presented in Table \ref{tab1}.
As an independent test of goodness-of-fit, we also compared the best-fit curves with a Nadaraya-Watson kernel density estimator from each ECDF \cite{Nadaraya64, Watson64}. Column (9) of Table 1 gives the probabilities, $Prob_K$, for these tests. Using the same method applied for best-fit parameter estimation, we determined the limits of the 95\% confidence interval of $q$ and $\sigma$. However, in some cases the lower limit of the confidence interval is imposed by the criterion for positivity of the probability given in equation (\ref{cutx}). In such cases, the limit is shown in brackets in columns (6) and (7) of Table \ref{tab1}. It is important to stress that although probability is lower than 50\% for most best-fit parameters, they are all rigorously within a 95\% confidence interval.

\section{RESULTS AND DISCUSSIONS}
Figure \ref{qxages} shows the variation in entropic index $q$ from rotation period distributions as a function of stellar ages. We can observe a tendency for an anticorrelation between $q$ and cluster age as reported in previous work by Silva et al. \cite{Silvaetal13}. Spearman's rank correlation test \cite{Hollander73} showed $\rho = -0.80$ and a probability of $p=0.002$ for the null hypothesis that the true $\rho$ is greater than 0. This result is very similar to that observed by Silva et al. \cite{Silvaetal13} for the time-dependence of entropic index $q$ estimated from the projected rotational velocity, $V\sin i$. It seems to indicate that the age-dependence of $q$ is not significantly affected by ambiguity due to the inclination of the stellar rotational axis. In fact, we found many similarities between our results and those reported by Silva et al. \cite{Silvaetal13}. First, the age for extensivity of the distributions, namely the age where the best fit line crosses line $q=1$, is $\log (age)=8.75 (\sim 600$ Myr), as shown in Figure \ref{qxages}. This age is quite similar to that of $\sim 700$ Myr found for $V\sin i$ data obtained by Silva et al. \cite{Silvaetal13}. Second, according to $q$ values given in Table 1, the minimum age for subextensivity ($q>1$) of rotation period distribution is around the age of NCG2516, namely 120 Myr, because it is the first cluster with an error bar crossing line $q=1$. This result is also similar to the age of $\sim170$ Myr found by Silva et al. \cite{Silvaetal13}. However, neither Table 1 nor Figure~\ref{qxages} presents a clear low limiting age for the superextensivity ($q<1$) of the distributions.

As mentioned in previous section, we used COCD ages because they provide a homogeneous scale of cluster ages, despite possible inaccuracies arising from the turn-off isochrone technique used by Kharchenko et al. \cite{Kharchenko05}. In fact, the ages given in COCD can be very different from those obtained in other studies (ex., compared with those shown in Table 1 by \cite{Gallet13}). In the present paper, such inaccuracies could affect, for example, estimation of the minimum ages for subextensivity and the age for extensivity of the distributions. However, it is worth noting that the error budget adopted here seems to be sufficient to account for the possible range of individual cluster ages. Accordingly, it can be stated that the minimum age for subextensivity falls within the range of the probable ages of NCG2516, namely between 60 and 270 Myr. On the other hand, if we consider the age of 1 Gyr for NGC6811, as determined by Janes et al. \cite{Janes13}, instead of 575 Myr, the age for the extensivity of rotation period distributions changes to only 560 Myr.

Figure \ref{qkxages} shows how the inverse of the median rotation period changes over cluster ages, as well as the nonextensive model proposed by de Freitas \& De Medeiros \cite{defreitas2013}. The best-fit lines for the data given in Table 1 are also displayed in Fig. \ref{qxages}. The unsaturated model with best-fit $q_K=2.5$ is illustrated by the black curve, the Skumanich Law with $q_K=3$ is represented by the blue curve, and the exponential decay law with $q_K=1$ is denoted by the red curve. The best-fit parameters for each curve are displayed in Table 2. According to Figure \ref{qkxages}, the model with different $q_K$ values can describe the changes in the period for clusters aged up to around 0.2 Gyr. However, exponential decay differs from the data when the clusters are older. On the other hand, the unsaturated model and the Skumanich law seem to be consistent with data for ages $\geq0.2$ Gyr, and there is no statistically significant distinction between the two models, as they have almost the same log-likelihood values ($LogLik$) (see Table 2). As showed de Freitas \& De Medeiros \cite{defreitas2013}, $q_{K}$ is given by the relationship $q_{K}=1+4aN/3$. In this respect, for the best-fit, $q_K=2.5$ indicates that for $a=1$ we have $N=1.1$, and for $a=2$, we have $N=0.6$. These values suggest that the magnetic field topology for the stars in our sample is slightly more complex than a purely radial field. According to Figure \ref{qkxages}, our sample stars never pass through the saturated regime, but rather pass directly to the unsaturated regime.  Table \ref{tab2} shows the best-fit values from the nonextensive model for the age-period relationship as displayed in Figure \ref{qkxages}. Table \ref{tab2} also demonstrates that the braking strength value $\lambda_{q}$ in the unsaturated regime is greater than in the saturated regime. The properties and significance of this parameter for the nonextensive model are defined in \cite{defreitas2013}.

As mentioned in previous section, the $q$-index quantifies the extension of the tail of the distribution. This denotes that a higher value of this index is related to an excess of stars with high rotation in the sample. According to Silva et al. \cite{Silvaetal13}, the slope of the correlation curve between $q$ and cluster ages is determined by $q_K$. In this respect, given the initial entropic index $q_{0}$ of the rotation period distribution for a stellar cluster, and the magnetic field topology for its stars, $q_{K}(a,N)$, we can estimate the age at which this distribution reaches the extensivity regime ($q=1$). At this age, the stars lose all their memory of initial angular momentum since we consider index $q$ as a memory parameter. In the main-sequence phase, magnetic braking is the main mechanism lowering stellar rotational velocities and therefore reducing the number of higher rotators over time, and this phenomenon can be scaled by parameters $q$ and $q_K$ from the nonextensitive formalism.

\section{CONCLUSIONS}
In this paper we analyzed the time-dependence of entropic index $q$ from rotation period data for 548 single main-sequence stars in 11 Open Clusters younger than 1 Gyr. We also analyzed the time evolution of rotation period distribution in the context of nonextensive formalism. As a result, we found that there is similar anti-correlation between $q$ from rotation period distribution to that found in previous works with $V\sin i$ data. We estimated an age for the extensivity regime of around 600 Myr, which is also very close to the age estimated from $V\sin i$ data. We interpret the extensivity regime as the states in which stars lose all their memory of initial angular momentum. Furthermore, we found that the variation in rotation period can be modeled by the nonextensive model for magnetic braking of stellar rotation with $q_{K}=2.5$, which corresponds to the unsaturated regime of the magnetic field. These results also indicate that the stars in our sample have a more complex magnetic field than a purely radial field. 

As a result, the values of $q$ reveal that the magnetic braking efficiency falls off with increasing age. In particular, in the extensive regime ($q=1$), this efficiency denotes the dividing line when the stars begin to lose their memory of initial angular momentum. This implies that the superextensive regime ($q<1$) represents the phase where the vast majority of the stars rotate very slowly. This correlation between the stellar rotation and the nonextensivity suggests that stellar rotation distribution is not simply a question of which mathematical formalism is applied, in fact, it depends on the statistical mechanics used. Moreover, in this  specific case, the nonextensive theory gives us a very good agreement with the observational data.

Finally, we would like to mention that we are conducting research based on large samples of old single field stars in the context of the present study in order to investigate the behavior of index $q$ as a scale parameter for the memory of initial angular momentum. The large number of rotation periods provided by \textit{Kepler} and CoRoT satellites also presents an excellent opportunity to test the nonextensitive model for magnetic braking of stellar rotation.

\acknowledgments
Research activity of the Stellar Board of the Federal University of Rio Grande do Norte (UFRN) is supported by continuous grants from CNPq and FAPERN Brazilian agency. B.B. Soares and J.R.P. Silva acknowledge financial support from the Programa Institutos Nacionais de Ci\^encia e Tecnologia (MCT-CNPq- Edital no. 015/2008).

\end{document}